# The Effect of Logarithmic Mesonic Potential on the Magnetic Catalysis in the Chiral Quark-Sigma Model

M. Abu-Shady

Department of Applied Mathematics, Faculty of Science, Menoufia University, Egypt

**Abstarct:** The chiral symmetry breaking in the presence of external magnetic field is studied in the framework of logarithmic quark-sigma model. The effective logarithmic mesonic potential is employed and is numerically solved in the mean-field approximation. We find that the chiral symmetry breaking enhances in comparison with the original sigma model. Two sets of parameterization are investigated in the present model. We find that an increasing coupling constant $g$ enhances the breaking symmetry while an increasing sigma mass $m_\sigma$ inhibits enhancing chiral broken vacuum state. A comparison with the Numbu-Jona-Lasinio model and the Schwinger-Dyson equation is discussed. We conclude that the logarithmic sigma model enhances the magnetic catalysis in comparison with the original sigma model and other models.

## 1-Introduction

The study of the influence of external magnetic fields on the fundamental properties of quantum chromodynamic (QCD) theory, confinement and dynamical chiral symmetry breaking is still a matter of great interest theoretical and experimental activities $[1-4]$. So for, most estimates have been carried out at vanishing chemical potential with the aid of effective theories such as the linear sigma model (LSM [5] and the Numbu-Jona-Lasinio model (NJL) [6, 7] in the mean field approximation.
The linear sigma model exhibits many of global symmetries of QCD theory. The model was originally introduced by Gell-Mann and Levy [8] with the purpose of describing pion-nucleon interactions. During the last years an impressive amount of work has been done with this model. The idea is to consider it as an effective low energy approach for QCD theory that the model has some aspects of QCD theory such as chiral symmetry. Thus, the model is successfully to describe the most of hadron properties at low energy such as in Refs. [5, 9, 10]. Some observables that are calculated in this model are conflict with experimental data. So researchers interest to modify this model to provide a good description of hadron properties such as in Refs. [11 − 14].

In Ref. [15], the authors have modified the linear sigma model by including the logarithmic mesonic potential and study its effect on the phase transition at finite temperature. In addition, the comparison with other models is done. On the same hand, the logarithmic sigma model successfully to describe the nucleon properties in low energy (zero temperature and density) [16].

To continue the investigation that started in Ref. [15]. In this work, we investigate the effect of external magnetic field on the chiral symmetry breaking in the framework of chiral logarithmic sigma model. In addition, a new parameterization for sigma mass and coupling constant on the behavior of the phase transition is studied.

This paper is organized as follows: The original sigma model is briefly presented in Sec. 2. Next, the effective logarithmic mesonic potential in the presence of external magnetic field is presented in Sec. 3. The results are discussed and are compared with other models in Sec. 4. Finally, the summary and conclusion are presented in Sec. 5.

## 2-The linear sigma model with original effective potential

The interactions of quarks via the exchange of $\sigma$ −and $\pi$- meson fields is given by the Lagrangian density [5] as follows:

$$L(r) = i\overline{\Psi}\partial_\mu \gamma^\mu \Psi + \frac{1}{2}(\partial_\mu \sigma \partial^\mu \sigma + \partial_\mu \boldsymbol{\pi} \cdot \partial^\mu \boldsymbol{\pi}) + g\overline{\Psi}(\sigma + i\gamma_5 \boldsymbol{\tau} \cdot \boldsymbol{\pi})\Psi - U_1(\sigma, \pi), \quad (1)$$

with

$$U_1(\sigma, \pi) = \frac{\lambda^2}{4}(\sigma^2 + \boldsymbol{\pi}^2 - v^2)^2 + m_\pi^2 f_\pi \sigma. \quad (2)$$

$U(\sigma, \boldsymbol{\pi})$ is the meson-meson interaction potential where $\Psi, \sigma$ and $\boldsymbol{\pi}$ are the quark, sigma, and pion fields, respectively. In the mean-field approximation, the meson fields are treated as time-independent classical fields. This means that we replace the power and products of the meson fields by corresponding powers and the products of their expectation values. The meson-meson interactions in Eq. (2) lead to hidden chiral $SU(2) \times SU(2)$ symmetry with $\sigma(r)$ taking on a vacuum expectation value

$$\langle \sigma \rangle = -f_\pi, \quad (3)$$

where $f_\pi = 93$ MeV is the pion decay constant. The final term in Eq. (2) included to explicitly breaking chiral symmetry. It leads to the partial conservation of axial-vector isospin current (PCAC). The parameters $\lambda^2, v^2$ can be expressed in terms of $f_\pi$, meson-masses as,

$$\lambda^2 = \frac{m_\sigma^2 - m_\pi^2}{2f_\pi^2}, \tag{4}$$

$$v^2 = f_\pi^2 - \frac{m_\pi^2}{\lambda^2}. \tag{5}$$

## 3-The effective logarithmic potential in the presence of magnetic field

In this section, the chiral logarithmic mesonic potential $U_2(\sigma, \pi)$ is employed. In Eq. (6), The effective logarithmic potential is extended to include the external magnetic field at zero temperature and density as follows,

$$U_{eff}(\sigma, \pi) = U_2(\sigma, \pi) + U_{Vaccum} + U_{Matter}, \tag{6}$$

where

$$U_2(\sigma, \pi) = -\lambda_1^2(\sigma^2 + \pi^2) + \lambda_2^2(\sigma^2 + \pi^2)^2 \log\left(\frac{\sigma^2 + \pi^2}{f_\pi^2}\right) + m_\pi^2 f_\pi \sigma, \tag{7}$$

In Eq. 7, the logarithmic potential satisfies the chiral symmetry when $m_\pi \to 0$ as well as in the standard potential in Eq. 2. Spontaneous chiral symmetry breaking gives a nonzero vacuum expectation for $\sigma$ and the explicit chiral symmetry breaking term in Eq. 6 gives the pion its mass.

$$\langle \sigma \rangle = -f_\pi. \tag{8}$$

Where

$$\lambda_1^2 = \frac{m_\sigma^2 - 7m_\pi^2}{12}, \tag{9}$$

$$\lambda_2^2 = \frac{m_\sigma^2 - m_\pi^2}{12f_\pi^2}. \tag{10}$$

For details, see Refs. [15, 16]. To include the external magnetic field in the present model, we follow Ref. [4] by including the pure fermionic vacuum contribution in the free potential energy. Since this model is renormalizable the usual procedure is to regularize divergent integrals using dimensional regularization and to subtract the ultra violet divergences. This procedure gives the following result

$$U_{Vaccum} = \frac{N_c N_f g^4}{(2\pi)^2}(\sigma^2 + \pi^2)^2 \left(\frac{3}{2} - \ln\left(\frac{g^2(\sigma^2 + \pi^2)}{\Lambda^2}\right)\right), \quad (11)$$

where $N_c = 3$ and $N_f = 2$ are color and flavor degrees of freedom. respectively, and $\Lambda$ is mass scale,

$$U_{Matter} = \frac{N_c}{2\pi^2} \sum_{f=u}^{d} (|q_f|B)^2 \left[\zeta^{(1,0)}(-1, x_f) - \frac{1}{2}(x_f^2 - x_f)\ln x_f + \frac{x_f^2}{4}\right] \quad (12)$$

In Eq. 12, we have used $x_f = \frac{g^2(\sigma^2 + \pi^2)}{(2|q_f|B)}$ and $\zeta^{(1,0)}(-1, x_f) = \frac{d\zeta(z, x_f)}{dz}\big|_{z=-1}$ representing the Riemann-Hurwitz function, and also $|q_f|$ is the absolute value of quark electric charge in external magnetic field with intense $B$. In Eq. 6, the part of effect of the finite temperature and chemical potential is not included in the present model which the present model focus on the study of magnetic catalysis at low energy due to the enhancement in the chiral symmetry breaking in the confinement state.

## Discussion of results

In this section, we study the effective potential of logarithmic sigma model. For this purpose, we numerically calculate the effective potential in Eq. (6). The parameters of the present model are the coupling constant $g$ and the sigma mass $m_\sigma$. The choice of free parameters of $g$ and $m_\sigma$ based on Ref. [4]. The parameters are usually chosen so that the chiral symmetry is spontaneously broken in the vacuum and the expectation values of the meson fields. In this work, we consider two different sets of parameters in order to get high and a low value for sigma mass. The first set is given by $\Lambda = 16.48 \text{ MeV}$ which yields $m_\pi = 138 \text{ MeV}$ and $m_\sigma = 600$ MeV. The second set as the first, yielding $m_\sigma = 400 \text{ MeV}$.

In Fig. 1, the effective potential is plotted as a function of sigma field for different values of magnetic field $(B)$. By ignoring $B$ independent one loop term in Eq. (6). We compare between the behavior of effective potential at zero and dense of magnetic field. We find that the effective potential increases at lower values of sigma field up to $\pm 75$ MeV. At higher values of $B$, the difference between the curves is strongly appeared. We note that the curve shifts to broken region of chiral symmetry by increasing $B$. Thus, the presence of intense magnetic fields leads to the chiral symmetry breaking enhances. This result is in agreement with the original sigma model in Ref. [17]. In comparison NJL model[18], Klevansky and Lemmer examined the chiral phase transition in the presence of electromagnetic field and they found the magnetic field enhances the chiral symmetry breaking. In comparison with the Schwinger-Dyson equation, Shushpanov and Smilga [19,20] studied the quark condensate in the presence of external magnetic field and found that the external magnetic field increases the quark condensate. In Refs. [21 → 23], the proper time method is applied to a four-fermion interaction model to study the influence of a magnetic field. It is shown that an external magnetic field has the effect of enhance chiral symmetry breaking. Therefore, the present results are in agreement with the original sigma model, the NJL model, and the Schwinger-Dyson equation.

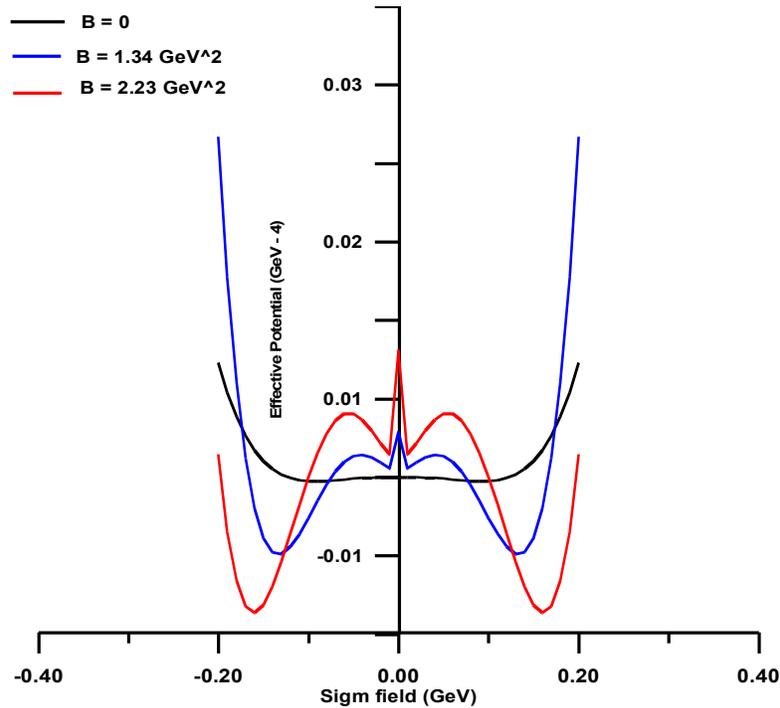

**Fig. 1:** The effective potential is plotted as a function of sigma field for sigma mass = 600 MeV and coupling constant g =5.38 for different value of magnetic field B.

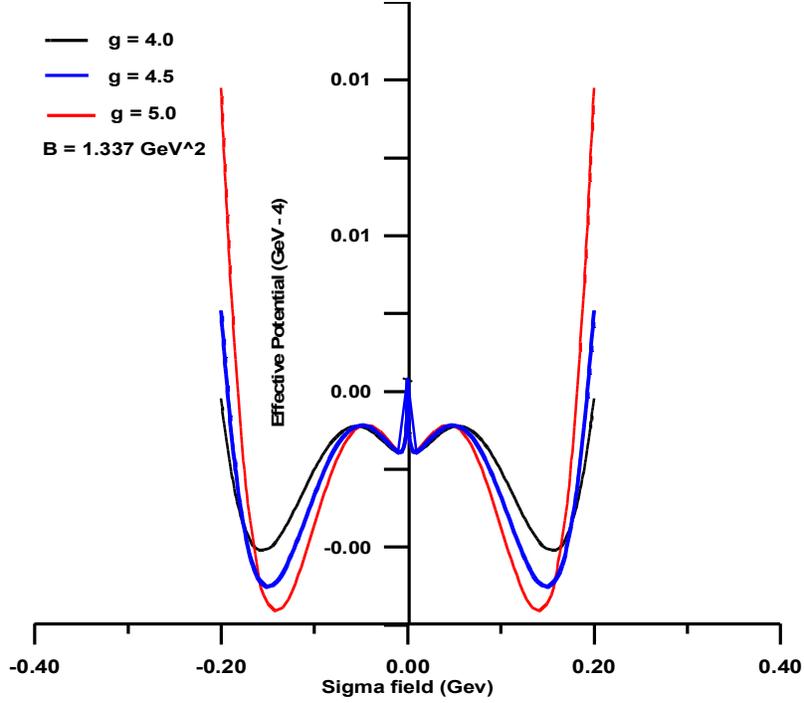

**Fig. 2.** The effective potential is plotted as a function of sigma filed at fixed magnetic field and different values of coupling constant g

It is important to discuss the effect of free parameters of the model on the chiral symmetry breaking. So, we select two sets of parameters. First set, the change of coupling constant with fixed sigma mass as in Fig. 2. The effective potential is plotted as a function of sigma field for different values of $g$ at dense of magnetic field $eB = 0.114 GeV^2$. We note that the behavior of effective potential is not sensitive for the change of sigma field up to $\pm 75$ MeV and then the effect of coupling constant $g$ strongly clarify when sigma field increases. We note that effective potential shifts to higher values by increasing $g$. Therefore, the effective potential is far from the chiral symmetry breaking by increasing $g$. So, the decreasing coupling constant $g$ enhanced the chiral symmetry breaking at dense of magnetic field. In the second set of parameters, we fixed the coupling constant $g$ with two values of sigma mass as in Fig. 3.

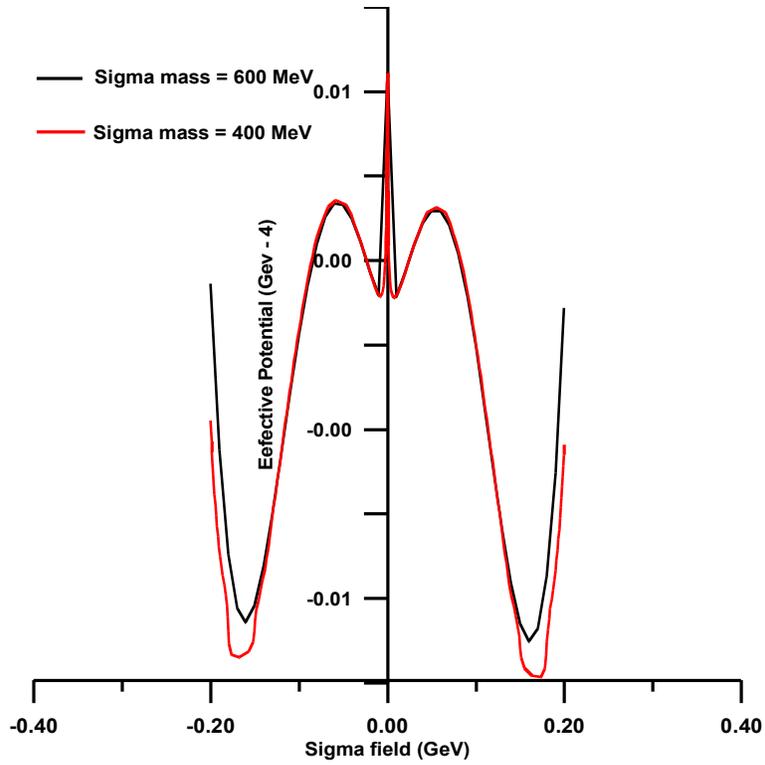

**Fig. 3.** The effective potential is plotted a function of sigma field at fixed value of magnetic field eB= 0.190 GeV$^2$ and pion mass = 138 MeV

We note that qualitative features of effective potential are remains unchanged, in which the effective potential is not sensitive up to $\simeq \pm 175$ MeV and then the potential shifts to lower values by increasing sigma mass. Therefore, the an increase in the sigma mass enhances the breaking of chiral symmetry at dense of magnetic field.

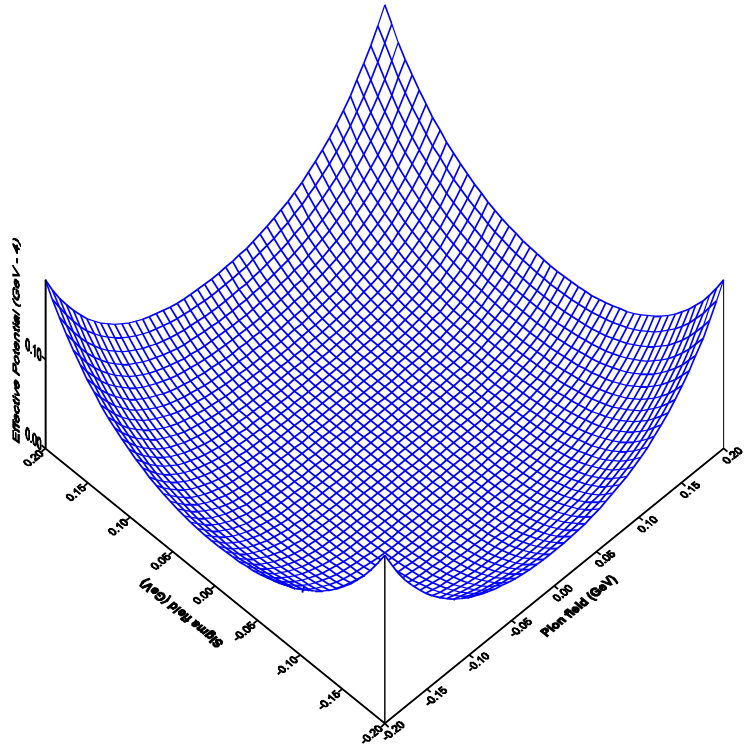

**Fig. 4:** The effective potential is plotted in the ($\sigma, \pi$) plane at B=0, sigma mass = 600 MeV, and g = 4.5

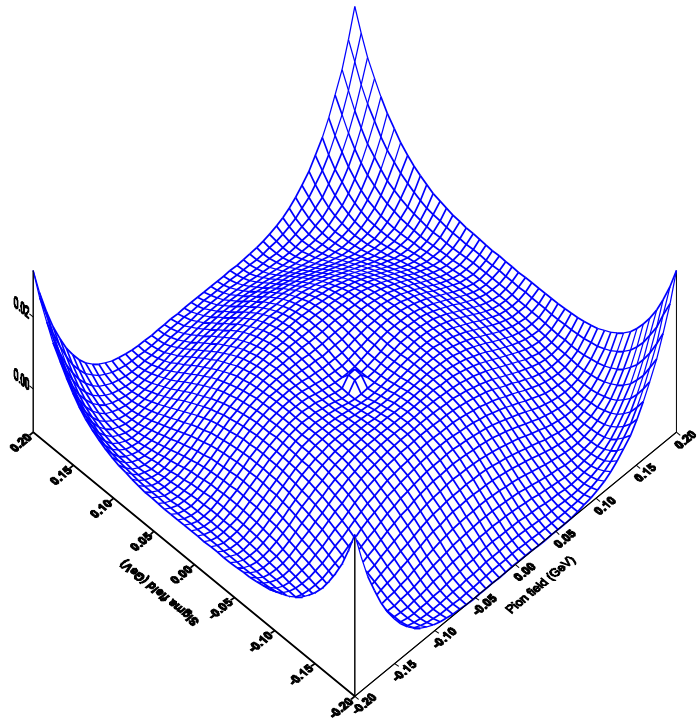

**Fig. 5**: The effective potential is plotted in the ($\sigma, \pi$) at eB=0.190 GeV$^2$, sigma mass = 600 MeV, and g = 4.5

In Figs. (4, 5), the effective of logarithmic potential is plotted in the ($\sigma, \pi$) plane at vanishing of magnetic field and strong magnetic field, respectively. In Fig. 4, we note that the potential is symmetric about point ($\sigma = 0, \pi = 0$). This symmetry is broken by increasing magnetic field as in Fig. 5.

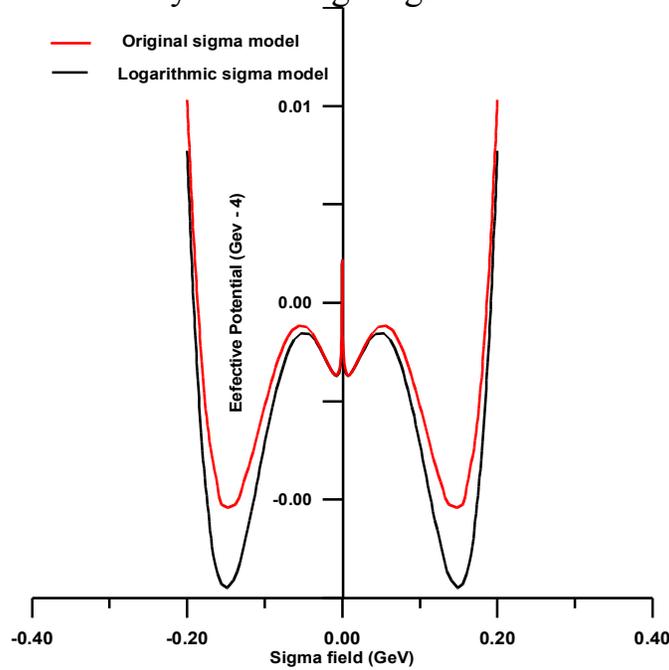

**Fig. 6.** The comparison between the chiral logarithmic model and the original sigma model g = 5.38, eB =0.114 GeV$^2$, and sigma mass = 600 MeV

In Fig. 5, we compare between the original sigma model and the logarithmic sigma model. We fixed all parameters in the two models to show realistic effect on the chiral symmetry breaking. Qualitative features are similar for the two model. We note that the difference between two model strongly appear when the sigma field increases. So, the effective logarithmic potential shifts to lower values in comparison with the original sigma model. Therefore, we conclude that the logarithmic sigma model enhances the chiral symmetry breaking.

## Summary and conclusion

In this work, we have employed the logarithmic potential to study the chiral symmetry breaking in the presence of magnetic field. So, novelty in this work, that the effect of the magnetic field is not investigated in the framework of logarithmic sigma model. This model applied only to calculate the hadron properties in low energy and chiral phase transition at finite temperature without considering magnetic field [15,16]. In addition, the effect of free parameters of the model is studied in the presence of strong magnetic field. A comparison with other models is presented, showing the obtained results are in the qualitative agreement with other models. Therefore, the logarithmic sigma model increases the magnetic catalysis which used for description of quantum particles as well as in the condensed matter. We hope to extended present model to finite temperature and density as a future work